\documentclass[twocolumn,showpacs,prb,superscriptaddress]{revtex4}
\usepackage{mathrsfs}
\usepackage{amsmath}
\usepackage{amsfonts}
\usepackage{bm}
\usepackage{amssymb}
\usepackage{graphicx}
\usepackage{dcolumn}
\usepackage{txfonts}
\usepackage{color}

\begin{document}


\title{Electrically-controllable RKKY interaction in semiconductor quantum
wires}
\author{Jia-Ji Zhu }
\author{Kai Chang }
\affiliation{SKLSM, Institute of Semiconductors, Chinese Academy of
Sciences, P. O. Box 912, Beijing 100083, China}
\author{Ren-Bao Liu }
\author{Hai-Qing Lin }
\affiliation{Department of Physics, The Chinese University of Hong
Kong, Shatin, New Territories, Hong Kong, China}

\begin{abstract}
We demonstrate in theory that it is possible to all-electrically manipulate
the RKKY interaction in a quasi-one-dimensional electron gas embedded in a
semiconductor heterostructure, in the presence of Rashba and Dresselhaus
spin-orbit interaction. In an undoped semiconductor quantum wire where
intermediate excitations are gapped, the interaction becomes the
short-ranged Bloembergen-Rowland super-exchange interaction. Owing to the
interplay of different types of spin-orbit interaction, the interaction can
be controlled to realize various spin models, e.g., isotropic and
anisotropic Heisenberg-like models, Ising-like models with additional
Dzyaloshinsky-Moriya terms, by tuning the external electric field and
designing the crystallographic directions. Such controllable interaction
forms a basis for quantum computing with localized spins and quantum matters
in spin lattices.
\end{abstract}

\maketitle

All-electrical quantum manipulation of the spin degree of freedom of
electrons and/or magnetic ions in semiconductors is a central issue in the
fields of spintronics and quantum information processing~\cite%
{Kane,Ohno,Tang,Molenkamp,nphys,Book}. Electron spins in semiconductors have
long coherence time and cost very low energy to flip~\cite{Awschalom}. These
features have obvious advantages for solid-state quantum information
processing where spins of electrons and magnetic ions have been proposed as
a candidate of qubits~\cite{Loss}. The Heisenberg-like exchange interaction
between two electrons confined in neighboring quantum dots can be controlled
electrically via changing the wavefunction overlap of the two electrons~\cite%
{Petta}. The distance between the neighboring dots is crucial for the
strength of the exchange interaction. In order to achieve lower accuracy
thresholds for quantum error correction, the implementation of coherent
long-distance interaction between two qubits is desirable. Optical field can
provide a practical way to realize the remote coupling between two local
spins via exchange interaction mediated by the cavity modes~\cite{Imamoglu},
and optically generated excitons and/or electrons~\cite{Sham,Pier}. Optical
control can be realized in femtosecond processes and made robust against
decoherence. The limitation of the spot size of laser beams, however,
hinders the integration of qubits under optical control. It is therefore
legitimate to design quantum gates based on electrically tunable remote
coupling between two spins.

The RKKY interaction is an indirect exchange interaction between localized
spins mediated by itinerant electrons in semiconductors or metals~\cite%
{RKKY,Imamura,Ziener}. The local spins can be magnetic ions or electron
spins in quantum dots~\cite{pnspin,nspin,Usaj,Qu}. A particularly
interesting system is quantum dots doped with single magnetic ions which
have strong \textit{s-d} exchange interaction with itinerant electrons~\cite%
{Kane}. Since the RKKY interaction is mediated by itinerant electrons, the
effects of spin-orbit interaction (SOI) are inevitable in conventional
zinc-blende semiconductors due to breaking of the crystal inversion
symmetry, i.e., Dresselhaus SOI (DSOI), and the structural symmetry, i.e.,
Rashba SOI (RSOI)~\cite{Meier,Nitta}. The SOI is one of the major sources of
spin decoherence and leads to an anisotropy in the relevant exchange
interaction~\cite{Imamura,Imamura2}. Such anisotropy in the RKKY interaction
arising from the SOIs is a resource to be exploited in this paper for
electrical control of various types of spin interactions, which is not
available in systems without SOIs.

In this work, we wish to draw attention to the possibility of creating spin
chains or lattices on semiconductor heterostructures. State-of-the-art
e-beam lithography makes it possible to fabricate such structures. We
demonstrate theoretically that the SOIs can be used to manipulate
electrically the symmetry type of the spin-spin interaction. The analytical
expression of the RKKY interaction shows the possibility of implementing
different quantum spin models by changing the strengths of the RSOI and
DSOI, e.g., isotropic and anisotropic Heisenberg models and Ising-like
model. A man-made spin lattice or chain mediated by this spin-spin
interaction would exhibit rich quantum phases.

\begin{figure}[t]
\includegraphics[width=\columnwidth]{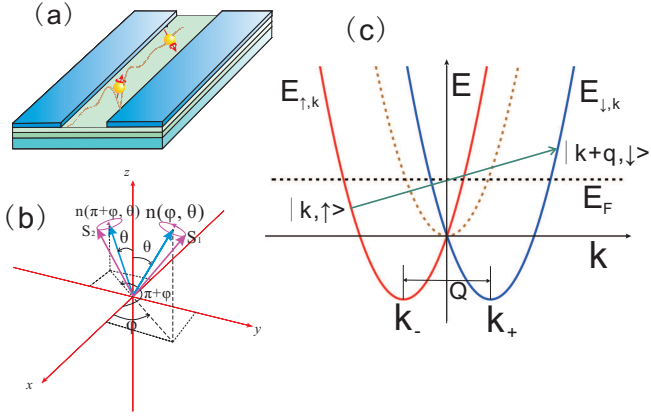}
\caption{(color online). (a) Schematic of two local spins in a quantum wire,
with electrically controllable RKKY interaction in the presence of RSOI and
DSOI. (b) Schematic of the angles $\protect\theta $ and $\protect\varphi $
in the twisted RKKY interaction. (c) Spin-split electron bands in the
presence of the SOI. The dashed curve denotes the band without the SOI. The
arrow indicates the spin-flip scattering.}
\label{fig:RKKY}
\end{figure}

First we consider two local spins $\mathbf{S}_{1}$ and $\mathbf{S}_{2}$
located at ${R}_{1}$ and ${R}_{2}$, mediated by electrons occupying the
lowest subband of a quantum wire in the presence of both the RSOI and DSOI
(see Fig.~\ref{fig:RKKY} (a)). The Hamiltonian of the system contains the
single-particle part $H_0$ and the s-d exchange interaction $H_1$ as
\begin{subequations}
\begin{align}
H =&H_{0}+H_{1},  \label{Hamiltonian} \\
H_{0} =&\sum_{k,\eta }E_{k\eta }c_{k\eta }^{\dag }c_{k,\eta}, \\
H_{1} =&J\underset{k,q,i,\eta,\eta^{\prime }}{\sum }e^{-iqR_i}c_{k+q,\eta^{%
\prime }}^{\dag }c_{k,\eta}{\boldsymbol{\sigma}}_{\eta^{\prime }\eta}\cdot{%
\mathbf{S}}_i,
\end{align}
where $c_{k \eta}$ annihilates an electron with quasi-momentum $k$ and spin $%
\eta$, ${\boldsymbol{\sigma}}$ denotes the Pauli matrices, and $J$ is the
strength of the s-d exchange interaction between itinerant electrons and the
local spins.

The non-interacting electron energy $E_{k\eta }$, determined by the
single-particle Hamiltonian $H_{0}=\hbar ^{2}k^{2}/2m^{\ast }+V(y)+H_{\text{%
SO}}$, is spin-dependent due to the SOI. Above $V(y)$ is the transverse
confining potential along the $y$ axis for electrons in the heterostructure,
$m^{\ast }$ is the electron effective mass, and $H_{\text{SO}}=\mathbf{B}_{%
\mathrm{eff}}\left( k\right)\cdot{\boldsymbol{\sigma }}$ is the SOI which is
equivalent to a momentum-dependent effective magnetic field $\mathbf{B}_{%
\mathrm{eff}}\left( k\right)$. The direction of the effective magnetic field
depends on the crystallographic plane and its strength is proportional to
the quasi-momentum in the linear SOI regime. For example, for typical
crystallographic planes $(001)$, $(110)$, and $(111)$, the effective field ${%
\mathbf{B}}_{\text{eff}}=k\left(\beta ,-\alpha,0\right) $, $k(0,-\alpha
,-\beta/2) $, and $k(0,-\alpha +2\beta /\sqrt{3},0)$ in turn (see Table~\ref%
{tab:H0thetaphi}), where $\alpha $ and $\beta $ are the strengths of the
RSOI and DSOI, respectively. If we choose the effective magnetic field as
the quantization direction $z^{\prime }$ for the spin, the energy bands are
split into two as (see Fig.~\ref{fig:RKKY} (c))
\end{subequations}
\begin{align}
E_{k,\pm}=\frac{\hbar^2}{2m^*}\left(k \pm\frac{Q}{2}\right)^2,
\end{align}
with the minima shifted in the momentum space by
\begin{align}
\pm Q/2\equiv \left(m^*/\hbar^2\right)B_{\text{eff}}(k)/k,  \label{shift}
\end{align}
for the spin parallel or anti-parallel to the $z^{\prime }$-axis,
respectively.

The RKKY interaction can be obtained using the Keldysh Green's function~\cite%
{schwabe}
\begin{eqnarray}
H_{1,2}^{RKKY} &=&-\frac{1}{\pi }\mathrm{Im}J^{2}\int_{-\infty
}^{\varepsilon _{F}}\mathrm{d}\varepsilon \mathrm{Tr}[(\mathbf{S}_{1}\cdot {%
\boldsymbol{\sigma }})G({R}_{12};\varepsilon +i0^{+})  \notag \\
&&\times (\mathbf{S}_{2}\cdot {\boldsymbol{\sigma }})G(-\mathbf{R}%
_{12};\varepsilon +i0^{+})],  \label{Keldysh}
\end{eqnarray}%
where $\varepsilon _{F}$ is the Fermi energy measured from the energy
minimum of the electron bands including the SOI, ${R}_{12}\equiv {R}_{1}-{R}%
_{2}$, and $\mathrm{Tr}$ means a partial trace over the spins of itinerant
electrons. With the spin quantization direction $z^{\prime }$ chosen along
the effective magnetic field ${\mathbf{B}}_{\text{eff}}$, the RKKY
interaction has the form
\begin{equation}
H_{1,2}^{RKKY}=F_{1}\left( q_{F}\left\vert R_{12}\right\vert \right) \left(
S_{1}^{z^{\prime }}S_{2}^{z^{\prime }}+\frac{1}{2}S_{1}^{^{\prime
}+}S_{2}^{^{\prime }-}e^{i2\theta }+\frac{1}{2}S_{1}^{^{\prime
}-}S_{2}^{^{\prime }+}e^{-i2\theta }\right) ,  \label{RKKY_SOI}
\end{equation}%
where $q_{F}=\sqrt{2m^{\ast }\varepsilon _{F}/\hbar ^{2}}$ is the Fermi
wavevector of each spin-split band measured from the band minimum, the phase
angle $\theta \equiv $ $Q\left\vert R_{12}\right\vert /2$, and the range
function is
\begin{equation}
F_{1}\left( q_{F}\left\vert R_{12}\right\vert \right) =\frac{4J^{2}}{\pi }%
\frac{m^{\ast }}{\hbar ^{2}}\left[ \mathrm{Si}\left( 2q_{F}\left\vert
R_{12}\right\vert \right) -\frac{\pi }{2}\right] ,  \label{rangefunction}
\end{equation}%
with $\mathrm{Si}(x)$ being the sine integral function. The physical
processes underlying different terms in the RKKY interaction is clearly
identified: The first term in the bracket of Eq.~(\ref{RKKY_SOI}) arises
from the spin-conserving scattering within each spin-split band and hence
has exactly the same form as in systems without the SOI~\cite{Yafet}. The
second and third terms correspond to the spin-flip scattering between
different spin-split bands and the phase factor $e^{\pm i2\theta }=e^{\pm
iQR_{12}}$ is the phase shift accumulated over $R_{12}$ by the extra
momentum transfer $Q$ that separates the minima of the two bands. The RKKY
interaction in Eq.~(\ref{RKKY_SOI}) is general for arbitrary
crystallographic planes and quantum wire orientations.

Back into the laboratory coordinate systems, the RKKY interaction is
transformed to
\begin{align}
H_{1,2}^{RKKY} = & F_{1}\left( q_{F}\left\vert R_{12}\right\vert \right) \{%
\mathbf{S}_{1}\cdot \mathbf{S}_{2}-2S_{1z}S_{2z}  \notag \\
&+2\left[ \mathbf{S}_{1}\cdot \mathbf{n}\left( \pi +\varphi ,\theta \right) %
\right] \left[ \mathbf{S}_{2}\cdot \mathbf{n}\left( \varphi ,\theta \right) %
\right] \},  \label{RKKY}
\end{align}%
where $\mathbf{n}\left( \varphi ,\theta \right) =\left( \cos \varphi \sin
\theta ,\sin \varphi \sin \theta ,\cos \theta \right) $. The angle $\theta $
and angle $\varphi $ are given in Table~\ref{tab:H0thetaphi} depend on the
crystallographic planes where the quantum wire is embedded. The angle $%
\theta $ describes how both the RSOI and DSOI twist the two local spins away
from the z axis and the angle $\varphi $ determines the in-plane twist of
the spin orientation (see Fig.~\ref{fig:RKKY} (b)). If the Dresselhaus
interaction is absent, our result is reduced to the previous work~\cite%
{Imamura}.

\begin{table*}[tbph]
\caption{The angles $\protect\theta $ and $\protect\varphi $\ describing the
effect of RSOI and DSOI in different crystallographic planes.}
\label{tab:H0thetaphi}%
\begin{ruledtabular}
\begin{tabular}{clcl}
Crystallographic planes&\multicolumn{1}{c}{$H_{0}$}&$\varphi$&\multicolumn{1}{c}{$\theta$}\\
\hline
(001)&$H_{0}^{(001)}=\frac{\hbar ^{2}k^{2}}{2m^{\ast}}+\alpha (\mathbf{k}%
\times \widehat{\mathbf{z}})\cdot {\boldsymbol \sigma}+\beta \left( \mathbf{%
k\cdot }\widehat{\mathbf{x}}\right) \left( \mathbf{\sigma \cdot }\widehat{%
\mathbf{x}}\right)$&$\arctan(\frac{\beta}{\alpha})$&$%
\frac{m^{\ast}}{\hbar^{2}}\sqrt{\alpha^{2}+\beta^{2}}|R_{12}|$\\
(110)&$H_{0}^{(110)}=\frac{\hbar ^{2}k^{2}}{2m^{\ast}}+\alpha (\mathbf{k}%
\times \widehat{\mathbf{z}})\cdot {\boldsymbol \sigma}-\frac{\beta}{2}\left( \mathbf{%
k\cdot }\widehat{\mathbf{x}}\right) \left( \mathbf{\sigma \cdot }\widehat{%
\mathbf{z}}\right)$&$\pi-\arctan(\frac{\beta}{2\alpha})$&$%
\frac{m^{\ast}}{\hbar^{2}}\sqrt{\alpha^{2}+\frac{\beta^{2}}{4}}|R_{12}|$\\
(111)&$H_{0}^{(111)}=\frac{\hbar ^{2}k^{2}}{2m^{\ast}}+(\alpha-\frac{2}{\sqrt{3}}%
\beta)(\mathbf{k}\times \widehat{\mathbf{z}})\cdot {\boldsymbol \sigma}%
$&$0$&$\frac{m^{\ast}}{\hbar^{2}}(\frac{2}{\sqrt{3}}\beta-\alpha)|R_{12}|$\\
\end{tabular}
\end{ruledtabular}
\end{table*}

The system with both the DSOI and RSOI has a great extent of
controllability, owing to the interplay of the two types of SOIs and the
sensitive dependence of DSOI on the crystallographic plane and the quantum
wire orientation. Eq.~(\ref{RKKY}) works for all but the $(110)$
crystallographic plane. For a quantum wire embedded in the heterostructure
grown along the $[110]$ direction, the RKKY interaction becomes $H\left(
\varepsilon _{F}\right) ^{RKKY}$ $=$ $F_{1}\left( q_{F}\left\vert
R_{12}\right\vert \right) $ $\left\{ \mathbf{S}_{1}\cdot \mathbf{S}%
_{2}-2S_{1x}S_{2x}+2\left[ \mathbf{S}_{1}\cdot \mathbf{n}\left( \pi +\varphi
,\theta \right) \right] \left[ \mathbf{S}_{2}\cdot \mathbf{n}\left( \varphi
,\theta \right) \right] \right\} $~\cite{note}. The RKKY interaction shows
isotropic behaviour at $\alpha $ $=$ $2\beta /\sqrt{3}$ when the quantum
wire is embedded on the $(111)$ crystallographic plane, because the RSOI and
DSOI have the same dependence on the in-plane momentum (see Table ~\ref%
{tab:H0thetaphi}). It means it is possible to switch on/off the SOIs \cite%
{Loss2}. When the quantum wire is embedded in arbitrary crystallographic
directions on $(001)$ plane, i.e. in the direction with an angle $\vartheta $
respect to the $[100]$ direction, the formalism of RKKY interaction remains
the same and the angle $\theta $ and angle $\varphi $ should be redefined as
$\theta \equiv m^{\ast }\sqrt{\alpha ^{2}+\beta ^{2}+2\alpha \beta \sin
2\vartheta }\left\vert R_{12}\right\vert /\hbar ^{2}$, $\varphi \equiv
\arctan \left[ \beta \cos 2\vartheta /\left( \alpha +\beta \sin 2\vartheta
\right) \right] $. Note that when the quantum wire is embedded along the $[%
\overline{1}10]$ direction on the $(001)$ plane, $\theta \propto \left\vert
\alpha -\beta \right\vert $ and $\varphi =0$, it would give us another way
to switch on/off the SOIs by tuning the strength of the external electric
field.

The most important difference between our result and the previous works~\cite%
{Sham,Pier} is that the interplay between the RSOI and DSOI offers us a new
way to control the spin-spin interaction. Due to the interplay of the DSOI
and RSOI, the RKKY interaction presents, in addition to the usual
Heisenberg-like exchange term, not only an Ising-type anisotropic term, but
also a twisted Dzyaloshinsky-Moriya (DM) -like term which twists the local
spins (see the third term in Eq. (\ref{RKKY})).

Tuning the parameters $\theta $ and $\varphi $, we can rotate the local
spins in a spin space (see Fig.~\ref{fig:RKKY} (b)), and construct various
kinds of quantum spin models. From Eq. (\ref{RKKY}), we can realize many
spin models by tuning the strengths of the SOIs (see Table~\ref%
{tab:spin_model}), for instance, isotropic Heisenberg-like model $%
F_{1}\left( q_{F}\left\vert R_{12}\right\vert \right) \mathbf{S}_{1}\cdot
\mathbf{S}_{2}$, the Ising-like model with an additional DM-like interaction
term $H\left( \varepsilon _{F}\right) ^{RKKY}$ $=$ $F_{1}\left(
q_{F}\left\vert R_{12}\right\vert \right) [S_{1x}S_{2x}\pm (\mathbf{S}%
_{1}\times \mathbf{S}_{2})_{x}]$ and the anisotropic Heisenberg model $%
H\left( \varepsilon _{F}\right) ^{RKKY}$ $=$ $F_{1}\left( \left\vert
R_{12}\right\vert \right) \left[ S_{1x}S_{2x}-S_{1y}S_{2y}-S_{1z}S_{2z}%
\right] $. In the long-range anisotropic Heisenberg model, along the
in-plane and the out-of-plane directions, the spin correlations, being
ferromagnetic or antiferromagnetic, are different.

The conditions mentioned in Table~\ref{tab:spin_model}, e.g., $\theta =k\pi $%
, are realizable in a narrow bandgap semiconductor InSb quantum well with $%
10 $ $\mathrm{nm}$ thickness at a specific perpendicular electric field $%
E\approx 50$ $\mathrm{kV}/\mathrm{cm}$. The SOI is strong in narrow bandgap
semiconductor quantum wells, e.g., HgCdTe QWs, in which $\alpha $ ranges
from $10^{-13}$ $\mathrm{eVm}\text{ to }10^{-10}$ $\mathrm{eVm}$ depending
on the external gate voltage, thickness of QW, and electron density~\cite%
{Kai}. Choosing a proper external electric field, one can realize the
switching between different spin models.

\begin{table*}[tbph]
\caption{Examples of various quantum spin models. $\pm $ corresponds to the
condition that the quantum wire is embedded in the $(001)$ plane but along
the $[100]$ $\left( +\right) $ or the $[010]$ $(-)$ direction, respectively.
}
\label{tab:spin_model}%
\begin{ruledtabular}
\begin{tabular}{c|llcc}
\multicolumn{1}{c}{Crystallographic
planes}\vline&\multicolumn{1}{c}{Spin
models}&\multicolumn{1}{c}{$H\left( \varepsilon
_{F}\right) ^{RKKY}$}&\multicolumn{1}{c}{$\theta$}&\multicolumn{1}{c}{$\varphi$}\\
\hline &Heisenberg model&$F_{1}\mathbf{S}_{1}\cdot \mathbf{S}_{2}$&$k\pi ,\left( k\in \mathbb{Z}\right)$&arbitrary values\\

(001)&Anisotropic Heisenberg model&$F_{1}\left[
S_{1x}S_{2x}-S_{1y}S_{2y}-S_{1z}S_{2z}\right]$&$\pi/2$&$\pi/2$\\

&Ising model with a DM term&$F_{1}
[S_{1x}S_{2x}\pm (\mathbf{S}_{1}\times \mathbf{S}_{2})_{x}]$&$\pi/4$&$\pi/2$\\

\hline &Heisenberg model&$F_{1}\mathbf{S}_{1}\cdot \mathbf{S}_{2}$&$k\pi ,\left( k\in \mathbb{Z}\right)$&arbitrary values\\

(110)&Anisotropic Heisenberg model&$F_{1}\left[-
S_{1x}S_{2x}-S_{1y}S_{2y}+S_{1z}S_{2z}\right]$&$\pi/2$&$\pi/2$\\

&Ising model with a DM term &$F_{1}[S_{1z}S_{2z}+(%
\mathbf{S}_{1}\times \mathbf{S}_{2})_{z}]$&$\pi/4$&$\pi/2 $\\

\hline &Heisenberg model&$F_{1}\mathbf{S}_{1}\cdot \mathbf{S}_{2}$&$k\pi ,\left( k\in \mathbb{Z}\right)$&$0$\\

(111)&Anisotropic Heisenberg model&$F_{1}\left[-
S_{1x}S_{2x}+S_{1y}S_{2y}-S_{1z}S_{2z}\right]$&$\pi/2$&$0$\\

&Ising model with a DM term &$F_{1}[S_{1y}S_{2y}+(%
\mathbf{S}_{1}\times \mathbf{S}_{2})_{y}]$&$\pi/4$&$0$\\
\end{tabular}
\end{ruledtabular}
\end{table*}

All-electrical two-qubit gates can be implemented with the RKKY interaction,
being either the Heisenberg-like interaction or the Ising-like interaction.
The controllability of the interaction symmetry in the SOI systems gives us
further flexibility of realizing various types of two-qubit gates such as
the $\sqrt{\mathrm{SWAP}}$ gate and the phase gate, either of which, plus
one-spin operations, constitute the complete set of gates for universal
quantum computing. In particular, the isotropic Heisenberg interaction can
be used for both one-qubit and two qubit gates~\cite{DiVincenzo}. For an
estimation of the operation rate, we notice that the exchange coupling $J$
can be tuned to $1$ \textrm{meV} by external electric fields, which
indicates that two-qubit gates with a picosecond cycle would be possible if
the electrical control can be done at that rate.

This approach of constructing electrical-controllable spin-spin interaction
outlined above can be extended to more complicated structures. Here we
propose that a single pair of spin qubits be replaced with an array of local
spins, i.e., a spin lattice or chain, which is defined on quantum wires
embedded in semiconductor heterostructures. Spin lattices are platforms of a
wealth of many-body physics and quantum phenomena such as quantum phase
transitions and may also be a computing resource such as in quantum
simulation of condensed matter systems. The RKKY interaction is a
long-ranged interaction since its asymptotic behavior $\lim_{R_{12}%
\rightarrow \infty }F_{1}\left( q_{F}R_{12}\right) \rightarrow \cos
(2q_{F}R_{12})/R_{12}$ that is inversely linear in distance $R_{12}$ with an
oscillation superimposed. In practice, precise positioning of spins for
realizing an artificial spin lattice proposed here is still a great
challenge. The long-range interaction would make the quantum physics richer
and more complicated. Such systems often manifest quantum phase transitions
governed by parameters such as the external field and concentration of
impurities. In order to realize a short-ranged spin-spin interaction, we
could use a one-dimensional intrinsic narrow bandgap semiconductor quantum
wire in which the virtual excitations between the valence and conduction
bands, in lieu of itinerant electrons in doped semiconductors, mediate the
interaction. The range function of the spin-spin interaction becomes $%
F_{1}\left( q_{F}R\right) \propto e^{-\lambda R}$ from Eq. (\ref{Hamiltonian}%
) utilizing the Keldysh Green's function. The interaction length $\lambda
\approx \hbar /\sqrt{2m^{\ast }\Delta }$, mostly determined by the electron
effective mass considering the large mass of the holes, can be tuned from $%
10 $ $\mathrm{nm}$ to infinity by adjusting the bandgap $\Delta $ of, e.g.,
a HgCdTe quantum well from $0.1$ $\mathrm{eV}$ to zero where a quantum phase
transition takes place~\cite{Kai2}. It provides us a new way to control the
range of the spin-spin interaction~\cite{Bloembergen}. Using virtual
excitations to mediate the spin-spin interaction also largely avoids the
fast optical decoherence~\cite{Sham}. In the spin lattice, one can also
control the spin-spin interaction spatially, and realize different spin
models in a spin lattice electrically, i.e., anisotropic Heisenberg model
and Ising-like model with an additional DM term. The DM-like term can induce
the interesting spiral phase in the spin chain in which the spins rotate
along the $\mathbf{S}_{i}\times \mathbf{S}_{i+1}$ axis. Using the twisted DM
term induced by the SOIs, one can use an electric field pulse, which
propagates along the spin chain, to generate a propagating spin wave along
the spin chain, and this spin wave excitation is actually a low power
consumption spin current since one only needs to flip the neighboring spins
without drifting of electrons.

However,\ the SOI in semiconductor low-dimensional electron gases is a
double-edged sword, since the spin relaxation is typically dominated by the
D'yakonov-Perel' (DP) mechanism~\cite{DP}, and is enhanced with increasing
the SOI.\ The spin decoherence induced by the SOIs is strongly suppressed in
this spin lattice due to the quasi-one-dimensional geometry of quantum
wires, since only one single point in the k-space satisfies the momentum and
energy conservation conditions for real excitations.

In summary, we propose all-electrical manipulation of the spin-spin
interaction via the RSOI and DSOI of electrons localized in quantum wires.
This RKKY interaction can be controlled in both magnitude and symmetry-tuned
heavily by adjusting the strength of SOIs, Fermi energy and crystallographic
planes, and display different types of spin-spin interactions. Both
isotropic and anisotropic Heisenberg models and Ising-like models with
additional DM terms could be realized. The anisotropy and twisted term in
the RKKY interaction caused by the SOIs can be removed by adjusting the
strength of SOIs. The parameters related to constructing the spin models can
be electrically controlled. Such in-situ controllability may be used for
observing quantum phase transitions in spin lattices without external
magnetic fields. The short-ranged spin-spin interaction can be realized
utilizing the virtual interband excitations in narrow bandgap semiconductors.

\begin{acknowledgments}
This work was supported by the NSFC Grant Nos. 60525405 and 10874175, and
the Knowledge Innovation Project of CAS, and Hong Kong RGC HKU 10/CRF/08. K.
C. appreciates the financial support from the C. N. Yang Fellowship and Hong
Kong RGC CUHK/401906.
\end{acknowledgments}


\begin{thebibliography}{99}
\bibitem{note} Here $\mathbf{n}\left( \varphi ,\theta \right) $ should be
redefined by exchanging the $x$ and $z$ component.
\end{thebibliography}
\end{document}